\documentstyle[psfig]{l-aa}
\begin{document}
\renewcommand{\topfraction}{1.}
\renewcommand{\bottomfraction}{1.}
\renewcommand{\textfraction}{0.}
\def\gsim{\vcenter{\hbox{$>$}\offinterlineskip\hbox{$\sim$}}}
\thesaurus{06(02.13.3; 08.03.4; 08.12.1; 08.13.2; 08.19.3; 11.13.1)}
\title{Discovery of H$_2$O maser emission from the red supergiant
       IRAS04553$-$6825 in the Large Magellanic Cloud}
\author{Jacco Th. van Loon\inst{1}, Peter te Lintel Hekkert\inst{2},
        Valent\'{\i}n Bujarrabal\inst{3}, Albert A. Zijlstra\inst{4}
        \and Lars-{\AA}ke Nyman\inst{5,6}}
\institute{Astronomical Institute ``Anton Pannekoek'', University of
           Amsterdam, Kruislaan 403, NL-1098 SJ Amsterdam,
           The Netherlands
      \and Australia Telescope National Facility, Parkes Observatory,
           P.O.Box 276, Parkes, NSW 2870, Australia
      \and Observatorio Astron\'{o}mico Nacional, Campus
           Universitario, Apartado 1143, E-28800 Alcal\'{a} de Henares
           (Madrid), Spain
      \and University of Manchester Institute of Science and
           Technology, P.O.Box 88, Manchester M60 1QD, United Kingdom
      \and European Southern Observatory, Casilla 19001, Santiago 19,
           Chile
      \and Onsala Space Observatory, S-439 92 Onsala, Sweden}
\date{Received 21 April 1998; accepted 25 May 1998}
\maketitle
\markboth{Jacco Th.\ van Loon et al.: circumstellar water maser in LMC}
         {Jacco Th.\ van Loon et al.: circumstellar water maser in LMC}
\begin{abstract}

We report the detection of 22 GHz H$_2$O maser emission from the red
supergiant IRAS04553$-$6825 in the Large Magellanic Cloud. It is the first
known source of circumstellar H$_2$O maser emission outside the Milky Way. The
measured flux density is comparable to that expected from scaling the galactic
red supergiant NML Cyg. The peak velocity agrees with the SiO maser peak
velocity.

A near-infrared spectrum indicates that IRAS04553$-$6825 has a typical LMC
metallicity. We argue that, possibly as a result of the low metallicity, the
H$_2$O emission probably occurs near or within the dust formation radius,
rather than further out as appears to be the case in NML Cyg and galactic
OH/IR stars. 

\end{abstract}

\section{Introduction}

Maser emission from evolved stars is associated with dusty, oxygen rich
circumstellar envelopes (CSEs), mainly around red supergiants (RSGs) and stars
at the tip of the Asymptotic Giant Branch (AGB) (Elitzur 1992; Habing 1996).
The strongest maser lines originate from OH (at 1612 MHz), H$_2$O (at 22 GHz),
and SiO (at 43 and 86 GHz), located progressively deeper into the CSE. Hence
these lines are used to trace the mass-loss history and velocity structure of
the CSE, through their photon fluxes and line profiles (Chapman \& Cohen 1986;
Lewis 1989, 1990). Maser lines are also excellent tracers of the stellar space
velocity and have been used to map the kinematics and mass distribution in our
Galaxy (e.g.\ Lindqvist et al.\ 1992; Sevenster et al.\ 1995).

Circumstellar OH and SiO maser emission has recently been detected in the
Large Magellanic Cloud (LMC) (Wood et al.\ 1986, 1992; van Loon et al.\ 1996,
1998a). Wood et al.\ (1992) argue that the outflow velocities in CSEs of OH/IR
stars in the LMC are a factor two lower than in corresponding Milky Way stars,
a difference they attribute to  less efficient driving of the outflow caused
by the lower metallicity in the LMC. However, Zijlstra et al.\ (1996) find the
difference to be only 20 to 30\%, from their analysis of the same data.
Moreover, the SiO maser emission from the LMC RSG IRAS04553$-$6825 (van Loon
et al.\ 1996) showed that in this case, the outflow velocity as derived from
the OH maser profile was underestimated by more than a factor two. The
relation between outflow velocity and metallicity is therefore still
uncertain.

In the LMC, SiO maser emission is generally too faint to be detected, with
IRAS04553$-$6825 the only successful detection to date. Stellar H$_2$O maser
emission has not been found outside the Galaxy before.

Here we present the discovery of the first extra-galactic source of
circumstellar H$_2$O maser emission --- viz.\ the RSG IRAS04553$-$6825 in the
LMC. We also present spectroscopic results that confirm the LMC metallicity of
its stellar photosphere. We then discuss the implications for the structure of
the CSE by comparing with the galactic red supergiant NML Cyg.

\section{Observations and Results}

\subsection{H$_2$O radio observations}

The observations were performed on August 19, 20 of 1997, using the 64~m radio
telescope at Parkes, Australia. We observed the $6_{16} \rightarrow 5_{23}$
rotational transition of ortho-H$_2$O at a rest frequency of 22.235 GHz, using
the 1.3~cm receiver plus autocorrelator backend. The 64~MHz band width and
1024 channels centred at $\sim$22.21~GHz yield a velocity coverage of
$\sim$860 km$\,$s$^{-1}$ at 0.84 km$\,$s$^{-1}\,$channel$^{-1}$. Using the
Dual Circular feed we simultaneously obtained spectra in left and right
circular polarization. The beam FWHM is $1.4^\prime$, the system temperature
is typically 110~K, and the conversion factor from antenna temperature to flux
density is 6.3 Jy$\,$K$^{-1}$. The nearby sky was measured every two minutes,
resulting in very flat baselines that required only a very shallow
second-order polynomial to be subtracted.

We observed the LMC RSG IRAS04553$-$6825 for six hours on-source integration.
No difference was found between the spectra obtained at either polarization,
which we then averaged. The final 22-GHz spectrum of IRAS04553$-$6825 is
presented in Fig.\ 1, with flux densities in mJy and heliocentric velocities
in km$\,$s$^{-1}$. The measured rms (1~$\sigma$) is only 5.5 mJy.

%
%
\begin{figure}[tb]
\centerline{\psfig{figure=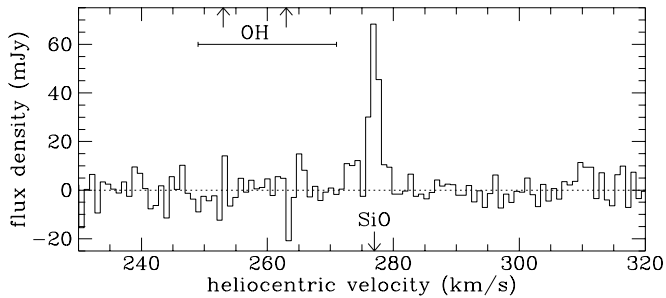,width=90mm}}
\caption[]{Discovery spectrum of the 22 GHz H$_2$O maser emission from
LMC red supergiant IRAS04553$-$6825. The velocities are heliocentric. Also
indicated are the peak velocities of the 86 GHz SiO and 1612 MHz OH masers
(arrows) and the total velocity extent of the 1612 MHz OH maser (bar)}
\end{figure}

We detected a single peak of H$_2$O maser emission with a peak flux density of
68~mJy (with 20\% calibration accuracy), corresponding to a 12~$\sigma$
detection. The peak is centred at a heliocentric velocity of 277
km$\,$s$^{-1}$, and has a FWHM of 1.7 km$\,$s$^{-1}$. The integrated flux of
the emission is 0.17 Jy$\,$km$\,$s$^{-1}$.

\subsection{CaII triplet echelle spectroscopy}

We used the 3.5~m New Technology Telescope (NTT) at the European Southern
Observatory (ESO) at La Silla, Chile, on January 7, 1996, with the ESO
Multi-Mode Instrument (EMMI), to obtain an echelle spectrum of
IRAS04553$-$6825. Grating \#14 and grism \#4 as cross disperser were used,
yielding a spectral coverage of 6000--9000 \AA. The slit width and length were
$2^{\prime\prime}$ and $4^{\prime\prime}$, respectively. The integration time
was one hour.

The data were reduced in the normal way using the Munich Interactive Data
Analysis Software (MIDAS) package. The wavelength calibration was done by
taking a ThAr lamp spectrum in conditions identical to the spectrum of
IRAS04553$-$6825. The measured spectral resolving power is $\sim4\times10^4$. 

The equivalent width of the Ca {\sc ii} triplet lines at 8498, 8542, and 8662
\AA\ measures the surface gravity in giants and supergiants, if the
metallicity is known. Reversely, if an estimate of the surface gravity is
available from knowledge of spectral type and luminosity, the Ca~{\sc ii}
lines can be used to infer the metallicity.

%
%
\begin{figure}[bt]
\centerline{\psfig{figure=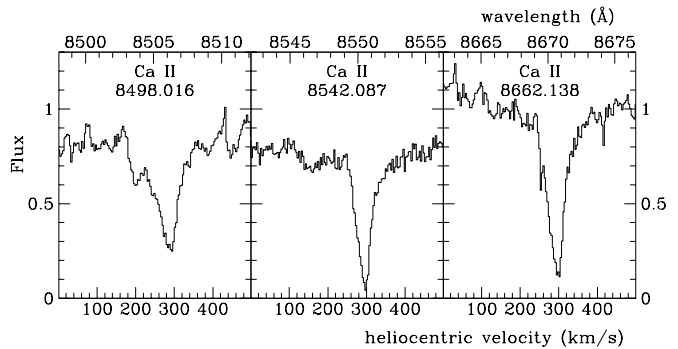,width=90mm}}
\caption[]{Echelle spectra of the Ca {\sc ii} triplet lines in
IRAS04553$-$6825}
\end{figure}

The echelle spectrum of IRAS04553$-$6825 around the Ca~{\sc ii} triplet is
presented in Fig.\ 2, on an arbitrary flux scale. The equivalent widths of the
two strongest components are measured as $W_{8542}=1.6\pm0.1$ \AA\ and
$W_{8662}=1.7\pm0.1$ \AA. Their sum is $W_{8542+8662}=3.3\pm0.2$ \AA.

The Ca~{\sc ii} absorption is maximum at a heliocentric velocity of $\sim$300
km$\,$s$^{-1}$, corresponding to a redshift with respect to the stellar
velocity of a little less than the outflow velocity in the outer parts of the
CSE. This is explained by scattering of the photospheric spectrum by an
extended, expanding dust shell. The measured equivalent widths of the Ca~{\sc
ii} lines are not affected by the scattering (see Romanik \& Leung 1981, and
references therein).

For IRAS04553$-$6825 we adopt a bolometric luminosity of $L\sim4\times10^5$
L$_\odot$, and an effective temperature between 2500 and 3000 K (spectral
type: M7 supergiant). The surface gravity must then be about $\log g\sim0$
cm$\,$s$^{-2}$. If we then compare the measured equivalent width of the
Ca~{\sc ii} triplet with the results from Garc\'{\i}a-Vargas et al.\ (1998),
we find that consistency is only achieved if the metallicity of
IRAS04553$-$6825 is a factor two to three lower than solar. This result is not
very sensitive to the values of the adopted stellar parameters, and hence we
conclude that IRAS04553$-$6825 has a typical LMC metallicity ($Z=0.008$).

\section{Discussion}

\subsection{Circumstellar H$_2$O masers in the LMC}

The detection of IRAS04553$-$6825 shows that H$_2$O masers in the LMC with
photon fluxes of $10^{44}$~s$^{-1}$ are detected at a 5~$\sigma$ level within
6 hours of on-source integration, with the current receiver at Parkes.
Typical observed H$_2$O maser photon fluxes from galactic sources are between
$10^{43}$~s$^{-1}$ (Lindqvist et al.\ 1990) and $10^{44-45}$~s$^{-1}$ (Nyman
et al.\ 1986) for OH/IR stars, $10^{41-44}$~s$^{-1}$ for irregular and
semiregular red variables with mass-loss rates $\sim10^{-7}$
M$_\odot\,$yr$^{-1}$ (Szymczak \& Engels 1997), $10^{43-44}$~s$^{-1}$ for Mira
variables (Benson \& Little-Marenin 1996; Yates et al.\ 1995), and
$10^{44-45}$~s$^{-1}$ for supergiants (Yates et al.\ 1995). This implies that
many of the mass-losing RSGs and AGB stars (i.e.\ IRAS point sources) in the
LMC are expected to have detectable H$_2$O maser emission.

\subsection{Introduction to IRAS04553$-$6825}

With a bolometric luminosity $M_{\rm bol}=-9.3$ mag and a spectral type M7,
IRAS04553$-$6825 was recognised to be the most luminous (very) red supergiant
in the LMC by Elias et al.\ (1986). The progenitor mass was estimated $M_{\rm
ZAMS}\sim50$ M$_\odot$ by Zijlstra et al.\ (1996). The current mass-loss rate
of $\dot{M}\sim$few $10^{-4}$ M$_\odot\,$yr$^{-1}$ was estimated from the
self-absorbed 10 $\mu$m silicate dust feature by Roche et al.\ (1993), who
argue that the lower than expected extinction in the optical indicates a disky
CSE.

IRAS04553$-$6825 exhibits strong OH 1612 MHz and weaker 1665 MHz mainline
emission (Wood et al.\ 1986, 1992). Comparison of the SiO maser peak velocity
and the OH maser emission profile revealed that OH is only observed from the
blue-shifted part of the shell. The outflow velocity is 27 km$\,$s$^{-1}$,
similar to Milky Way RSGs. (Wood et al.\ derived 10 km$\,$s$^{-1}$ as is
suggested by the OH maser emission profile alone.) It is possible that the
red-shifted emission is much weaker than the blue-shifted and therefore below
the detection limit. Stronger blue-shifted emission is expected if the maser
amplifies emission from within the shell. Alternatively, an asymmetric and
structured OH profile can also arise from bipolar outflow (Chapman 1988; te
Lintel Hekkert et al.\ 1988).

\subsection{Similarities between IRAS04553$-$6825 and NML Cyg}

IRAS04553$-$6825 is remarkably similar to NML Cyg, a well known RSG in the
Milky Way (Johnson 1967; Diamond et al.\ 1984; Richards et al.\ 1996; Monnier
et al.\ 1997), in terms of luminosity, progenitor mass, spectral type,
pulsation period, mass-loss rate, 10 $\mu$m feature profile, outflow velocity,
CSE geometry, OH maser emission profile and peak flux density (scaled to a
common distance of 1 kpc), and SiO maser photon flux. The properties of the
two stars are compared in Table 1. We compiled the infared photometry from the
data provided in Gezari et al.\ (1993), and use $F_{8 {\mu}m}({\rm zero
mag})=2\times10^{-12}$ W$\,$m$^{-2}\,\mu$m$^{-1}=42.7$~Jy. The H$_2$O
properties (NML Cyg: Yates et al.\ 1995) are also included in Table 1.

The pump efficiency of SiO masers is known to be related to the pulsation
amplitude (Alcolea et al.\ 1990): Mira variables and semi-regular variables
with visual amplitudes exceeding 2.5 mag always reach the maximum efficiency,
while variables with smaller amplitudes are usually less efficient. We
recalculated the pump efficiency of the SiO masers $F_{\rm SiO}/F_{8 {\mu}m}$.
The pumping in IRAS04553$-$6825 is twice as efficient as in NML Cyg, but a
factor $\sim4$ less than in Mira variables. The SiO maser emission of
IRAS04553$-$6825 has a larger ratio $F_{\rm int}/F_{\rm max}$ than NML Cyg (at
a common resolution of 0.7 km$\,$s$^{-1}$), and a few times as large as Mira
variables (Alcolea et al.\ 1990; van Loon et al.\ 1996).

The near-infrared (JHKL) amplitudes are 0.4 mag for NML Cyg (Gezari et al.\
1993) and 0.3 mag for IRAS04553$-$6825 (Wood et al.\ 1992). For NML Cyg,
Kholopov et al.\ (1985) list a magnitude range of 11.19 to 12.54 mag in R, and
17.0 to 18.0 mag in V. We have sparse optical photometry of IRAS04553--6825 in
the period 1994--1997, suggesting an amplitude in V of $\sim2$ mag (smaller in
R). We conclude that the pulsation properties of NML Cyg and IRAS04553$-$6825
are similar, with possibly IRAS04553$-$6825 closer resembling Mira variables.

From the similar properties of both objects we predicted to detect 22 GHz
H$_2$O maser emission from IRAS04553$-$6825 at a level of $\sim60$ mJy (van
Loon 1998b). The measured peak flux density is in excellent agreement with
this prediction. 

%
%
\begin{table}[tb]
\caption[]{Comparison between the properties of the red supergiants
NML Cyg in the Milky Way and IRAS04553$-$6825 in the LMC.}
\begin{tabular}{l|ll}
\hline\hline
                    & NML Cyg                  & IRAS04553$-$6825        \\
\hline
Bolo.\ Luminosity   & $5\times10^5$ L$_\odot$  & $4\times10^5$ L$_\odot$ \\
Distance            & 2 kpc                    & 50 kpc                  \\
Progenitor Mass     & $\sim$50 M$_\odot$       & $\sim$50 M$_\odot$      \\
Spectral Type       & M6                       & M7                      \\
Pulsation Period    & 940 days                 & 930 days                \\
Mass-loss Rate      & 1.8$\times10^{-4}$ M$_\odot$ yr$^{-1}$ & $\sim$5$\times10^{-4}$ M$_\odot$ yr$^{-1}$ \\
10 $\mu$m Silicate  & emission+absorption      & emission+absorption     \\
$(K-L)$ colour      & 2.4 mag                  & 1.9 mag                 \\
$(H-K)$ colour      & 2.0 mag                  & 1.2 mag                 \\
Outflow Velocity    & 28 km s$^{-1}$           & 27 km s$^{-1}$          \\
CSE Geometry        & disk + bipolar           & disk + bipolar          \\
OH Profile          & blue asymm./duplicit     & blue asymm./duplicit    \\
OH Peak (1 kpc)     & 1.1 kJy                  & 1.0 kJy                 \\
OH Photon Flux     & $4\times10^{45}$ s$^{-1}$ & $3.7\times10^{45}$ s$^{-1}$ \\
SiO Profile         & peak + broad comp.       & peak + broad comp.      \\
SiO Peak (1 kpc)    & 144 Jy                   & 270 Jy                  \\
SiO Photon Flux  & $0.6\times10^{45}$ s$^{-1}$ & $2.6\times10^{45}$ s$^{-1}$ \\
$F_{\rm SiO}/F_{8 {\mu}m}$ & $\frac{1}{90}$    & $\frac{1}{43}$          \\
$F_{\rm int}/F_{\rm max}$(SiO) & 7.3 km s$^{-1}$ & 15.7 km s$^{-1}$      \\
H$_2$O Profile      & blue-shifted maximum     & stellar velocity        \\
H$_2$O Peak (1 kpc) & 160 Jy                   & 170 Jy                  \\
H$_2$O Photon Flux & $2.6\times10^{44}$ s$^{-1}$ & $2.6\times10^{44}$ s$^{-1}$ \\
\hline
\end{tabular}
\end{table}

\subsection{Location of the H$_2$O masers}

A striking difference between the H$_2$O masers of IRAS04553$-$6825 and NML
Cyg is that in IRAS04553$-$6825 the emission peaks at the stellar restframe
velocity, whereas in NML Cyg the line profile resembles the OH maser line
profile.

Interferometric observations of the masers around NML Cyg indicate that its
H$_2$O masers are amplified radially and that they are located in the dusty
part of the CSE, where radiation pressure on the grains accelerates the
outflowing matter (Richards et al.\ 1996). Indeed, in NML Cyg the strongest
emission occurs at a blueshift of $\sim20$ km$\,$s$^{-1}$.

The peak of H$_2$O maser emission from IRAS04553$-$6825 falls within $\sim1$
km$\,$s$^{-1}$ of the stellar velocity (as inferred from the SiO velocity),
and all emission is detected within a range of 8 km$\,$s$^{-1}$. This is
similar to the situation in Mira variables, indicating low outflow velocities
in the H$_2$O masing region of the CSE (Yates et al.\ 1995). In contrast, NML
Cyg is similar to the OH/IR stars.

Cooke \& Elitzur (1985) derive a scaling relation for estimating the inner
radius $r_q$ of the H$_2$O masing region, below which the density is higher
than $\sim10^9$~cm$^{-3}$ and the maser is quenched. This was confirmed by
observations (Yates \& Cohen 1994). The collision rate is dominated by
collisions with H$_2$ and does therefore not depend on metallicity: we
therefore expect that the same relation will hold in the LMC. With an outflow
velocity at $r_q$ of 0.8 km$\,$s$^{-1}$ (half the FWHM of the H$_2$O maser
peak), and an average particle mass of $2\times10^{-24}$~g, we estimate
$r_q\sim4\times10^{15}$~cm, or $\sim20$ times the stellar radius. This is
outside the region where the SiO maser occurs (Humphreys et al.\ 1996).

We speculate that in IRAS04553$-$6825 the H$_2$O maser originates near the
dust formation radius, where the logarithmic velocity gradient is very high
($d(\ln(v))/d(\ln(r))\gsim1$) and tangential amplification dominates over
radial amplification. Part of the emission may also originate from inside the
dust formation radius where outflow velocities are low. Both cases could
explain the narrow H$_2$O peak coinciding with the tangentially amplified SiO
maser peak (see also Engels et al.\ 1997). The H$_2$O masers in NML Cyg are
located outside of the dust formation region, where the outflow velocities are
already $\gsim15$ km$\,$s$^{-1}$ (Richards et al.\ 1996).

In this model, either the maser occurs closer to the star in IRAS04553$-$6825,
or dust formation takes place further out (as in galactic Miras as compared to
galactic OH/IR stars). The first possibility appears less likely if the inner
radius is set by collisional quenching of the population inversion. It would
be interesting to test whether at lower metallicity the dust formation and the
acceleration of the outflow occurs at larger radii because of the relative
lack of refractory elements: in this case the difference between NML Cyg and
IRAS04553$-$6825 could be a generic difference between similar stars in the
Milky Way and the LMC. 

\acknowledgements{We would like to thank Drs.\ Marcus Price and Ian Stewart
for help with the observations at Parkes. We also thank Dr.\ Roland Gredel for
help with the NTT observations at La Silla, that were performed in Director's
Discretionary Time. We thank Dr.\ Anita Richards for helpful discussions, and
an anonymous referee for her/his remarks that helped improve the paper. This
research was partly supported by the NWO under Pionier Grant 600-78-333. O
jacco agradece ao anjinho Joana por ser para ele como ele queria ser para
ela.}

\end{document}